\documentclass[aps,prl,floatfix,superscriptaddress,
twocolumn,
footinbib,longbibliography]{revtex4-2}
\usepackage{amssymb}
\usepackage{amsmath}
\usepackage{amsfonts}
\usepackage{appendix}
\usepackage{bm}
\usepackage[caption=false,subrefformat=parens]{subfig}
\usepackage{graphicx}
\usepackage{epsfig}
\usepackage{epstopdf}
\usepackage{balance}
\usepackage[dvipsnames]{xcolor}
\usepackage{calc}
\usepackage{natbib}
\usepackage[colorlinks,
linkcolor=blue,
anchorcolor=blue,
citecolor=blue,
urlcolor=blue]{hyperref}
\usepackage{lipsum}
\usepackage{braket}
\usepackage{subfig}
\usepackage{hyperref}
\usepackage{verbatim}
\usepackage{cleveref}

\begin{document}

\title{Nonlinear optical thermodynamics from a van der Waals–type mean-field theory}

\author{Meng Lian}
\affiliation{School of Physics and Wuhan National High Magnetic Field Center, Huazhong University of Science and Technology, Wuhan 430074,  China}
\author{Zhongfei Xiong}
\affiliation{Institut national de la recherche scientifique, Centre Énergie Matériaux Télécommunications (INRS-EMT), 1650 Blvd. Lionel-Boulet, Varennes, QC J3X 1S2, Canada}
\author{Yuntian Chen}
\email{yuntian@hust.edu.cn}
\affiliation{School of Optical and Electronic Information, Huazhong University of Science and Technology, Wuhan 430074, China}
\author{Jing-Tao L{\"u}}     
\email{jtlu@hust.edu.cn}
\affiliation{School of Physics and Wuhan National High Magnetic Field Center, Huazhong University of Science and Technology, Wuhan 430074,  China}

\begin{abstract}
Optical thermodynamics offers a distinctive framework for understanding complex phenomena in multimode systems, yet standard ideal-gas-like  formulation neglects the effect of nonlinear interaction on thermodynamic quantities, significantly restricting its range of validity. Here, we overcome this limitation by developing a mean-field thermodynamic theory that incorporates the nonlinear renormalization of the mode spectrum. The resulting nonlinear equation of state, analogous to that of the van der Waals for gases, enables the prediction of power-dependent mode localization and the description of optical cooling and heating in photonic Joule–Thomson expansion. Our work establishes a unified thermodynamic perspective on the nonlinear control and transport of optical waves.
\end{abstract}

\maketitle

Multimode optical systems exhibit a wealth of physical phenomena, and thermodynamic theory is gradually emerging as a powerful framework for understanding their complex behavior,
circumventing the need to explicitly model the microscopic system dynamics\cite{wu_thermodynamic_2019, parto_thermodynamic_2019, makris_statistical_2020, wright_physics_2022, Mangini:24}. It has offered fresh insights into a wide range of optical effects, including beam self-cleaning \cite{krupa_spatial_2017, liu_kerr_2016, niang_spatial_2019}, nonequilibrium evolution \cite{shi_controlling_2021, PhysRevLett.132.193802, PhysRevLett.133.116303, PhysRevResearch.7.013084}, topological effect\cite{PhysRevLett.126.073901, jung_thermal_2022, Mančić_2023, Wong2025}, optical condensation \cite{PhysRevLett.125.244101, Bloch2022, PhysRevA.110.063530}, while also providing new opportunities for the design of high-performance optical devices \cite{Barsi2009, doi:10.1126/science.aao0831, Richardson2013, Wehbi:22}.

By treating optical power and `internal energy' as thermodynamic invariants, standard linear optical thermodynamic theory models highly multimode optical systems as ideal gases.  Wherein, the only role of nonlinearity is to drive the system toward thermodynamic equilibrium. While valid at low power input \cite{wu_thermodynamic_2019, pourbeyram_direct_2022}, linear theory fails to capture, even qualitatively, genuine nonlinear phenomena such as soliton formation \cite{LEDERER20081},  phase transition \cite{PhysRevX.10.031024}, and optical Joule–Thomson (J-T) expansion \cite{Kirsch2025, Pyrialakos:25, Dinani2025, Pyrialakos2025}. 
Extending thermodynamic theory to incorporate inter-modal interaction could thus substantially broaden its range of applicability, paving the way of nonlinear light manipulation guided by thermodynamic principles.

\begin{figure}
\centering
\includegraphics[width=0.49\textwidth]{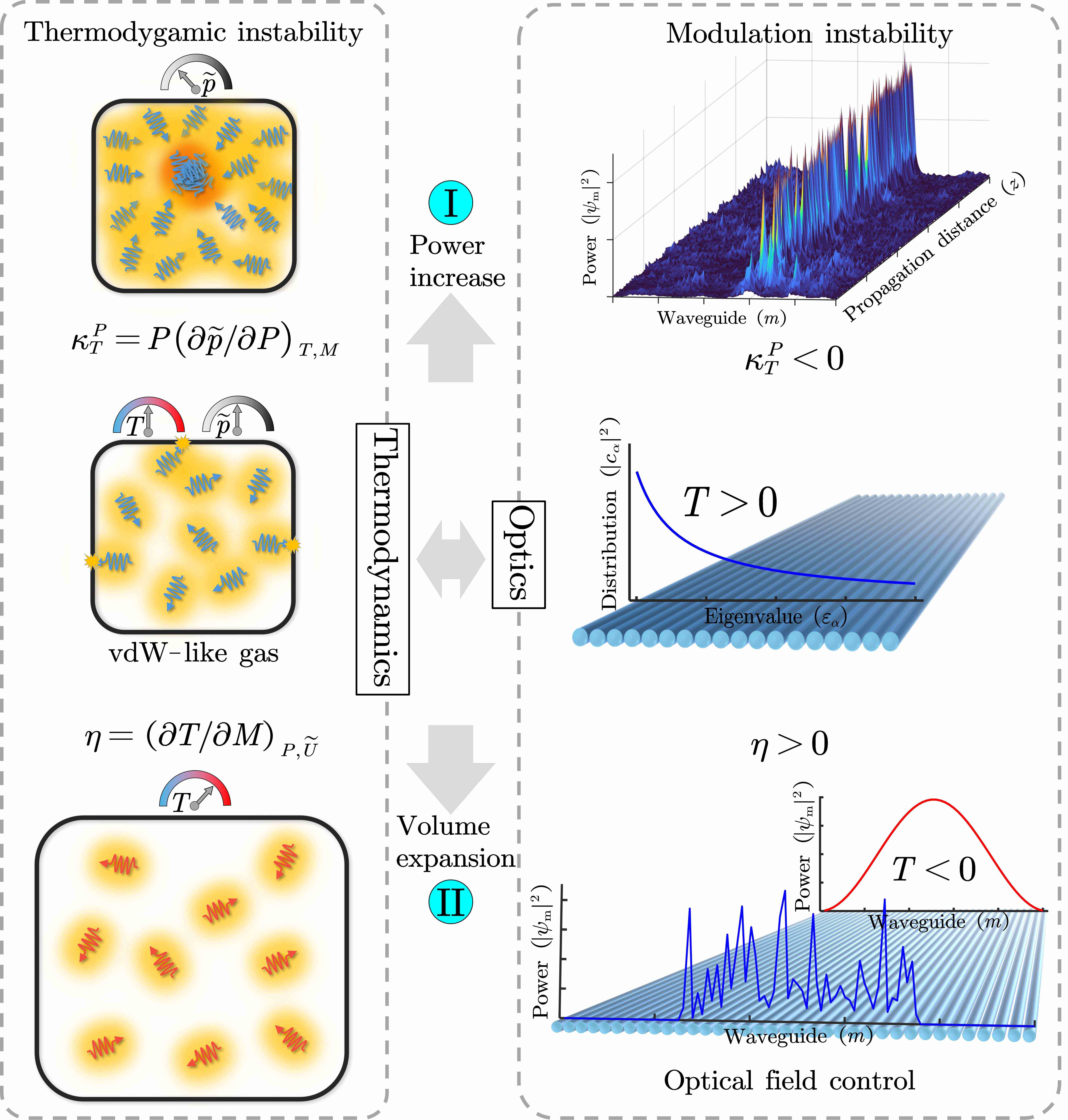}
\caption{{\textbf{Nonlinear thermodynamic theory together with response functions enables understanding of nonlinear optical phenomena from a distinct viewpoint.}} 
The left column shows the thermodynamic processes of a vdW-like optical gas, while the right column shows the corresponding nonlinear optical processes. A vdW-like optical gas model (middle left) describes the multimode optical waveguide array. The mode occupation follows the R-J distribution in equilibrium (middle right). (I) Increasing the optical power at fixed temperature and mode volume results in thermodynamic instability described by a negative value of the isothermal power elastic modulus $\kappa^P_T$ (top left).  The gas undergoes a gas–liquid-like phase separation, corresponding to modulation instability in optics (top right). (II) Increasing the mode volume at fixed power and internal energy (optical J-T expansion) results in temperature change, characterized by the optical J-T expansion coefficient $\eta$ (bottom left). For $\eta>0$, the gas temperature increases from positive to negative value, giving rise to Rayleigh-Jeans condensation at negative temperature. This can be used for optical field control (bottom right).
}
     \label{fig:1}
\end{figure}

In the thermodynamics of classical gases, one solution to such a limitation is to introduce an effective mean-field potential that accounts for inter-particle interaction. The van der Waals (vdW) theory is a prototypical example\cite{van1873over,callen1993}.
%
%
Here, we adopt a similar idea. Working in the basis of eigenmodes, we decompose the total Hamiltonian into two parts: the mean-field Hamiltonian $H_\text{mf}$, which includes the linear Hamiltonian together with the renormalization of mode frequencies due to nonlinear interaction, and the interaction Hamiltonian $H_{\text{int}}=H-H_\text{mf}$, which describes interaction among the renormalized modes. Based on this decomposition, we develop a nonlinear optical thermodynamic theory, whose equilibrium state still follows the Rayleigh–Jeans (R-J) distribution, but with renormalized thermodynamic quantities. 
The resulting equation of state (EoS) becomes a higher-order function of the mode volume and optical power. Unstable solution of the EoS in the high-power regime is directly linked to soliton formation. When applied to optical J–T expansion, the theory enables a thermodynamic understanding of optical heating or cooling, accompanied by energy transfer between the linear and nonlinear parts of the internal energy (Fig.~\ref{fig:1}). The framework is universally applicable to systems of different dimensions, thereby providing a thermodynamic foundation for the nonlinear control of optical systems.

 
{\bf Model --} Propagation of the optical modes along the waveguide array is described by 
the discrete nonlinear Schr\"odinger equation (DNSE)\cite{LEDERER20081,agrawal_applications_2001}
\begin{equation} 
i\frac{d\psi _m}{dz}=-\kappa \sum_{\langle n\rangle}{\psi _n}+\chi \left| \psi _m \right|^2\psi _m=0,
  \label{equ:DNES}
\end{equation}
where dimensionless parameters $\psi_m$, $\kappa$, $\chi$ represent the complex wave amplitude, the nearest neighbor coupling among waveguides and Kerr-type nonlinear coefficient, respectively. The positive/negative sign of $\chi$ indicates repulsive/attractive nonlinear interaction.
Here the coordinate $z$ along waveguide plays the role of time in the standard Schr\"odinger equation.

Equation~(\ref{equ:DNES}) corresponds to an effective Hamiltonian
\begin{equation} 
H=H_\text{L}+H_\text{NL}=-\sum_{\langle m,n\rangle}{\kappa _{mn}\psi _{m}^{*}\psi _n}+\sum_m{\frac{\chi}{2}|\psi _m|^4},
\label{equ:H_t}
\end{equation}
For weak nonlinear interaction, we can diagonalize the linear term and obtain the corresponding eigen mode (supermode) with propagating constant $\beta_\alpha$ and corresponding eigen vector $\phi_\alpha$. With the total number of modes $M$, the eigen energy defined by the negative of $\beta_\alpha$ as $\varepsilon_\alpha = -\beta_\alpha$ with $\beta _1\ge \beta _2\ge \cdots \ge \beta _M$. The system thus has a bounded spectrum between $[\varepsilon_1, \varepsilon_M]$, and  
can effectively achieve negative optical temperature \cite{Baudin2023, marques2023observation, wu_entropic_2020, PhysRev.103.20}.
For a given state $\psi$, the modal occupancy is obtained by projection onto each eigen mode
$\left|c_\alpha\right|^2=\left| \langle \phi _\alpha | \psi \rangle \right|^2$.

{\bf Nonlinear optical thermodynamic theory --} 
We can write the mean-field Hamiltonian in the eigen mode basis
\begin{equation} 
H_\text{mf}=\sum_{\alpha}{\varepsilon _{\alpha}|c_{\alpha}|^2}+\chi \sum_{\alpha ,\beta}{\Gamma_{\alpha \beta \beta \alpha}|c_{\beta}|^2}|c_{\alpha}|^2,
\label{equ:H0}
\end{equation}
where 
\begin{equation} 
\Gamma _{\alpha \beta \gamma \delta}=\sum_m{\phi _{\alpha}^{*}\left( m \right) \phi _{\beta}^{*}\left( m \right) \phi _{\gamma}\left( m \right) \phi _{\delta}\left( m \right)}
\end{equation}
is the four-wave mixing coefficient. Different from the linear theory, here we retain part of the nonlinear contribution, which gives rise to a shift of the eigen energy  $\tilde{\varepsilon}_{\alpha}=\varepsilon _{\alpha}+2\chi \sum_{\beta}{\Gamma _{\alpha \beta \beta \alpha}|c_{\beta}|^2}$, without inducing any mode exchange [Supplementary Material (SM)\footnote{See Supplemental Material, which includes Refs.~\cite{shi_controlling_2021,wu_thermodynamic_2019,PhysRevE.89.022921,PhysRevLett.132.193802,parto_thermodynamic_2019}, for details on the process of formula derivation, and additional numerical results.}, S1]. The remaining term $H_{\text{int}}$ is responsible for the scattering among modes.  

\begin{figure*}
	\centering
   \includegraphics[width=\textwidth]{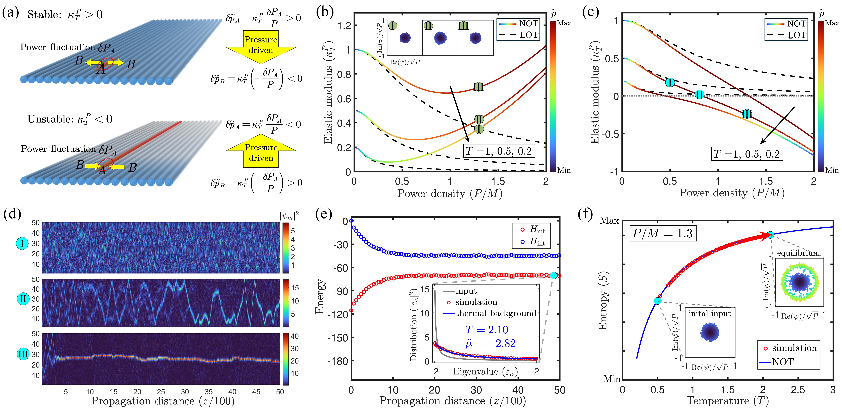}
\caption{{\textbf{Thermodynamic prediction of instability in 1D waveguide array at large input power.}} (a) Schematic of the waveguide array with $M=50$, $\kappa=1$. The initial input satisfies a R-J distribution. When a local power fluctuation $\delta P_A>0$ occurs in region A, the power in region B decreases by $\delta P_A$. When the isothermal power elastic modulus $\kappa^P_T>0$, the pressure $\tilde{p}$ in region A increases while that in region B decreases, such that power is transferred from the high-pressure region A back to the low-pressure region B and the fluctuation is suppressed. When $\kappa^P_T<0$, the $\tilde{p}$ in region A decreases whereas that in region B increases, causing more power transferred into region A and amplifying the fluctuation. This thermodynamic instability results in power localization. The yellow arrows indicate the direction of power transfer. (b) The $\kappa^P _T$ as a function of input power density $P/M$ for $\chi=0.1$. The solid and dashed lines are results from the nonlinear (NOT) and linear (LOT) theories, respectively. Each solid line is color coded to represent normalized pressure at each temperature. The inset shows the complex-$\psi$ space distribution calculated from the DNSE at $P/M=1.3$, with the color indicating the relative magnitude of optical power $|\psi _m|^2/P$. (c) Similar results as (b), but for $\chi=-0.1$. $\kappa_T^{P}$ becomes negative at high power density. (d) Evolution of the optical power distribution $|\psi _m|^2$ along $z$ at $P/M=$0.5 (I), 0.8 (II), and 1.3 (III), as marked in (c). One ensemble is selected randomly from numerical simulation of the DNSE in each plot. (e) Energy stored in the mean-field ($H_\text{mf}$) and the interaction Hamiltonian ($H_\text{int}$) during evolution along $z$ at $P/M=$1.3. The inset shows the corresponding equilibrium distribution. (f) Optical entropy as a function of temperature at $P/M=1.3$, with a red arrow indicating evolution from the initial to the final state. The inset shows the complex-$\psi$ space distribution obtained from ensemble average of 500 runs. }
\label{fig:soliton}
\end{figure*}

The system evolves with conserved internal energy
$\widetilde{U}=\langle H_\text{mf} \rangle $
and optical power
$P=\sum_m^M{\left| \psi _m \right|^2}=\sum_{\alpha}^M{\left| c_{\alpha} \right|^2}$. 
Maximizing the optical entropy 
$S=\sum_{\alpha}^M{\ln \left| c_\alpha \right|^2}$,  
the mode occupation can be shown to follow the classical R-J distribution (SM, S2-A\cite{Note1})\cite{PhysRevE.89.022921} 
\begin{equation} 
|c_{\alpha}|_{eq}^{2}=\frac{T}{\tilde{\varepsilon}_{\alpha}-\tilde{\mu}},
\label{equ:RJ}
\end{equation}
where $T$ and $\tilde{\mu}$ are the optical temperature and chemical potential, respectively. We use the  tilded symbols to represent quantities defined in the nonlinear theory, which are different from the linear version.  
Equation~(\ref{equ:RJ}) is also the equilibrium solution of the kinetic equation derived from the mean-field Hamiltonian $H_\text{mf}$ with the frequency‑shift correction (SM, S1\cite{Note1}).

We adopt the approximate relation $\sum_{\beta}{\Gamma _{\alpha \beta \beta \alpha}|c_{\beta}|_\text{eq}^{2}} = P/M$ to write the frequency renormalization in terms of macroscopic quantities (SM, S2-B\cite{Note1}) \cite{PhysRevE.89.022921, PhysRevLett.132.193802}. In so doing, compared to the linear theory, the temperature $T$ remains unaffected, while the chemical potential and the internal energy are modified by the nonlinear corrections (SM, S3\cite{Note1})
\begin{align}
\label{equ:muz}
\tilde{\mu}&=\mu +2\chi P/M, \\
\widetilde{U}&=U_L+U_{NL}=\sum_{\alpha}{\varepsilon _{\alpha}\left| c_{\alpha} \right|^2}+\chi ^{\left( M \right)}{P^2}/{M}.
\label{equ:Uz}
 \end{align}
Here, $\chi ^{\left( M \right)}$ is $\chi/2$ for $M=1$ and is $\chi$ otherwise. The EoS is then obtained from Eqs.~(\ref{equ:muz}-\ref{equ:Uz})
\begin{equation}
\widetilde{U}-\tilde{\mu}P=MT-\chi ^{\left( M \right)}{P^2}/{M}.
\label{equ:u1}
\end{equation}
From this EoS, we can show that the optical entropy satisfies the fundamental thermodynamic relation (SM, S2-C\cite{Note1})
\begin{equation}
dS=\frac{1}{T}d\widetilde{U}-\frac{\tilde{\mu}}{T}dP+\frac{\tilde{p}}{T}dM,
\end{equation}
with the optical thermodynamic pressure $\tilde{p}$
\begin{equation}
\tilde{p} = \hat{p}+\chi \frac{P^2}{M^2}=\frac{ST}{M}-T+\chi \frac{P^2}{M^2}.
\label{equ:p}
\end{equation}
In addition to pressure defined in the linear theory  $\hat{p}$, it acquires a nonlinear correction which closely parallels the vdW gas pressure correction.

The inclusion of nonlinear correction enables a consistent description of thermalization between subsystems with different nonlinear interactions, a limitation of the linear theory (SM, S4\cite{Note1}). More importantly, together with the different types of thermodynamic response functions, the theory enables an understanding of nonlinear optical processes from a thermodynamic point of view, which we illustrate in the following.

\begin{figure*}
\centering
\includegraphics[width=\linewidth]{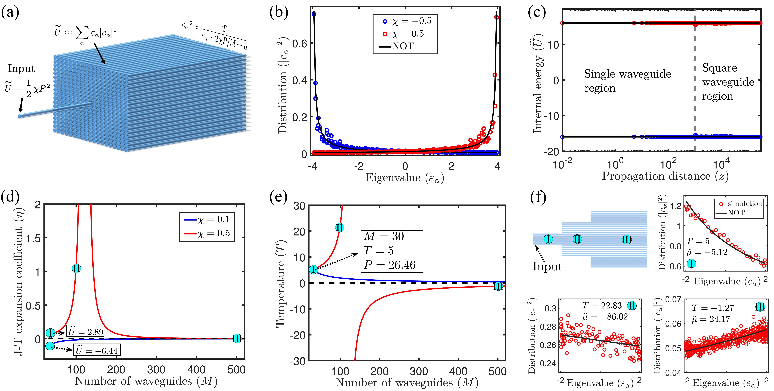}
\caption{{\textbf{Cooling and heating in optical J-T expansion described by the nonlinear optical thermodynamic theory.}} Parameter: $\kappa=1$. Lines: theory; Dots: numerics. (a) Schematic of a structure realizing J-T expansion: a single waveguide couples into the center of a 19 $\times$ 19 square array. (b) The equilibrium modal distribution in the 19 $\times$ 19 array for input $P=8$, $\chi=\pm 0.5$. (c) The internal energy $\widetilde{U}$ is conserved during the expansion. (d) Variation of the optical J–T response coefficient $\eta$ with $M$ for a 1D homogenous waveguide array. Initial condition:  $M=30$, $T=5$, $P=26.46$. (e) Corresponding temperature evolution for the case in (d). (f) Schematic of a three-step expansion and corresponding modal distribution for $\chi=0.5$. The array size increases from the initial $M=30$ (I) to $M=100$ (II), and then to $M=500$ (III).  }
\label{fig:J-T}
\end{figure*}

{\bf vdW–like optical gas instability --}  
For an optical waveguide array with a fixed “volume” $M$, the isothermal power elastic modulus $\kappa^P _T=P\left( \partial \tilde{p}/\partial P \right) _{T,M}$ quantifies the response of optical thermodynamic pressure to changes in optical power. It can be shown to be equivalent to the isothermal volumetric elastic modulus $\kappa _{T}^{M}=-M\left( \partial \tilde{p}/\partial M \right) _{T,P}$. Using Eq.~(\ref{equ:p}) and the Maxwell relation, it is expressed as (SM, S5\cite{Note1})
\begin{equation}
\kappa^P _T=-\frac{PT}{M}\left( \frac{\partial {\mu}}{\partial T} \right) _{P,M}+2\chi \frac{P^2}{M^2},
\end{equation}
The two terms on the right-hand side represent the linear and nonlinear contributions, respectively. The sign of the linear term is the same as the optical temperature (SM, S5\cite{Note1}). 
Inclusion of nonlinear correction could change this picture. In the case of $\kappa^P _T$, the system is thermodynamically unstable: a small power fluctuation in the system can be amplified [Fig.~\ref{fig:soliton}(a) and SM S6\cite{Note1}]. 
%
%
Figure \ref{fig:soliton}(b-c) shows the dependence of $\kappa^P_T$ on the power density $P/M$ for a 1D homogeneous array [Fig.~\ref{fig:soliton}(a)] at positive temperature.
More results at negative temperature and for other lattice  types are shown in Supplementary Figs.~S2-S5\cite{Note1}, demonstrating the universality of the theory. 
For a positive $\chi=0.1$ [Fig.~\ref{fig:soliton}(b)], the repulsive interaction increases $\kappa^P _T$. 
However, for negative value $\chi=-0.1$, $\kappa^P _T$ becomes negative as the power density increases [Fig.~\ref{fig:soliton}(c)], indicating an anomalous pressure response, i.e., larger optical power results in smaller pressure, rendering the system unstable.

To gain further insights, we have solved the DNSE numerically. Distribution of optical power during propagation for three representative power densities $P/M$ = 0.5 (I), 0.8 (II), and 1.3 (III) is depicted in Fig.~\ref{fig:soliton}(d). They correspond to positive (I), near zero (II) and negative (III) $\kappa^P_T$, respectively, as marked in Fig.~\ref{fig:soliton}(c). Drastically different evolution patterns are found for the three cases. 
In case I, the system evolves according to the initial equilibrium distribution, which macroscopically manifests as a random, thermalized spreading across the entire array. This is similar to the behavior for positive $\chi$ (not shown). In case II, the competition between nonlinearity and diffraction becomes pronounced, giving rise to intermittent localization within the waveguide array. This is the operating region for nonlinear optical funneling \cite{Dinani2025}. Further increasing the power density (III) results in strong localization with robust, soliton‑like evolution trajectories. This corresponds to the modulation instability in nonlinear optics. At this stage [Fig.~\ref{fig:soliton}(e)],  a large portion of the internal energy is transferred from $H_\text{mf}$ to $H_{\text{int}}$. The initial input decomposes into a thermal background, which obeys the R-J distribution [Fig.~\ref{fig:soliton}(e), inset], and a localized component. This is highly analogous to gas–liquid phase separation in the unstable region of vdW gases. The localized component here corresponds to the liquid phase where nonlinear inter-modal interaction dominates, while the thermal background corresponds to the gas phase.  
The nonlinear theory therefore makes a connection between the unstable thermodynamic response and the optical modulation instability\cite{LEDERER20081, PhysRevA.105.033529}. 

When localization occurs, obtaining the final state requires taking $H_{\text{int}}$ into account\cite{Rasmussen2000, PhysRevE.69.016618} in order to maximize the background entropy. According to Fig.~\ref{fig:soliton}(f), the ideal maximum entropy state corresponds to $T\rightarrow \infty $.  In simulations, however, dynamical effects such as nonlinearity and lattice pinning limit further energy transfer \cite{PhysRevE.69.016618}, causing the $H_{\text{int}}$ to saturate. The temperature eventually stabilizes at $T=2.10$, and the complex-$\psi$ space evolves from a dense thermal cloud into a high-power ring contributed by localized waveguides and a remaining thermal cloud formed by the other waveguides [Fig.~\ref{fig:soliton}(f), inset].

{\bf  Cooling and heating in optical Joule-Thomson expansion --}
Similar to real gases, ``volumetric'' expansion of the optical system is accompanied by temperature change due to energy transfer between the linear and nonlinear parts of the Hamiltonian. The optical J–T expansion, which corresponds to a sudden increase of $M$, has been demonstrated to offer new opportunities for light routing using nonlinear processes based on thermodynamic principle\cite{Kirsch2025,Pyrialakos:25,Dinani2025,Pyrialakos2025}. As an example, for the structure shown in Fig.~\ref{fig:J-T}(a),
the nonlinear theory captures the key factor that the initial input is entirely determined by the nonlinear energy. When $M$ is suddenly increased, it is transferred into the linear part, thereby controlling the temperature.
%
%
%
Figure~\ref{fig:J-T}(b) shows the final modal distribution for $\chi=\pm 0.5$ with an input optical power $P=8$. The system equilibrates at $T=\mp 0.045$,  respectively. During these processes, $\widetilde{U}$ is conserved [Fig.~\ref{fig:J-T}(c)]. The theoretical prediction agrees with the numerical results. 

We can define another thermodynamic quantity, the optical J–T expansion coefficient, to characterize the temperature change during a self-similar change of $M$ (SM, S7\cite{Note1})
\begin{equation}
\eta=\left(\frac{\partial T}{\partial M} \right) _{P,\widetilde{U}} =-\frac{T}{M}+\chi \frac{P^2}{M^2}\left( \frac{\partial T}{\partial U_L} \right) _{P,M}.
\end{equation}
The first and second terms correspond to the linear and nonlinear contributions, respectively. The nonlinear term can change the simple monotonic $M$ dependence of the linear one. Specifically, for $T>0$, $\chi>0$, the two terms have opposite sign and compete with each other. We consider two typical cases in the simplest 1D homogenous waveguide array [Fig.~\ref{fig:J-T}(d-f)]. For $\chi=0.1$ (blue), the linear term dominates. Although the transfer of nonlinear energy into the linear part suppresses the cooling, we still have $\eta<0$. For larger $\chi=0.5$ (red), the nonlinear theory gives results quite opposite to the linear one. In this case, $\eta$ changes sign and becomes positive. Consequently, expansion results in heating instead of cooling predicted by the linear theory. As shown in Fig.~\ref{fig:J-T}(e),  the temperature increases sharply, becomes negative for $M\sim 120$, and eventually heats up toward $T\rightarrow 0^-$. 

To validate the theoretical prediction,   
Fig.~\ref{fig:J-T}(f) illustrates benchmark of the theory (black lines) with exact numerical calculations (red dots).  
The 1D waveguide array undergoes two expansion $M=30(\text{I}) \to 100 (\text{II}) \to 500 (\text{III})$. This corresponds to heating at positive temperature (I$\to$II) and the transition to a negative-temperature state (II$\to$III), respectively. The agreement with numerical results (see also Supplementary Fig.~S6\cite{Note1} for $T<0$) 
demonstrates that the nonlinear theory can provide a theoretical basis for temperature control using nonlinear effect in waveguide arrays.

In summary, we have developed an optical thermodynamic theory, taking into account the nonlinear interaction within a mean field approximation, akin to the van der Waals theory for gases. The modified equation of state enables the prediction of soliton formation, cooling/heating in optical J-T expansion. All these can be described by corresponding thermodynamic response functions, thus providing a unified and generic thermodynamic framework to understand complex nonlinear processes in multimode optical systems.

\emph{Note added --} During the preparation of this work, we became aware of a related work \cite{kabat2026}, which focuses on 1D case in the strong nonlinear regime.

{\bf Acknowledgments --} This work was supported by the National Key Research and Development Program of China (Grant No. 2022YFA1402400) and the National Natural Science Foundation of China (Grant No. 22273029).

\bibliography{lian}

\clearpage        
\onecolumngrid    

\setcounter{figure}{0}
\setcounter{equation}{0}
\renewcommand\thefigure {S\arabic{figure}}
\renewcommand\theequation {S\arabic{equation}}

\section*{Supplementary}  

\section{S1. Kinetic equation}

The effective Hamiltonian of  a nonlinear optical waveguide array is given by
\begin{equation} 
H=H_\text{L}+H_\text{NL}=-\sum_{m,n}{\kappa _{mn}\psi _{m}^{*}\psi _n}+\sum_m{\frac{\chi}{2}|\psi _m|^4},
\label{equ:H_t}
\end{equation}
where $\psi _{m}$ denotes the complex amplitude of the optical field in the $m$-th  waveguide, which forms a pair of canonical conjugate variables with $i\psi _{m}^{*}$. Evaluating $\partial \psi _m/\partial z=\partial H/\partial \left( i\psi _{m}^{*} \right) $, we obtain the discrete nonlinear Schrödinger equation governing the system dynamics
\begin{equation} 
i\frac{d\psi _m}{dz}=-\sum_n{\kappa _{mn}\psi _n}+\chi \left| \psi _m \right|^2\psi _m.
\label{equ:DNS}
\end{equation}
Here $z$ is the coordinate along the propagation direction, playing the role of time in normal Schr\"odinger equation. 
Considering the weakly nonlinear regime, the eigen states of the system are determined by the linear part $H_L$ of the Hamiltonian. The complex amplitude of the optical field at $z$ in the $m$-th waveguide is given by
\begin{equation} 
\psi _m\left( z \right) =\sum_{\alpha}^M{c_{\alpha}\left( z \right) \phi _{\alpha m}},
\label{equ:eigenmode}
\end{equation}
where $\phi _{\alpha}$ denotes the $\alpha$-th eigen mode of $H_L$, and $c_{\alpha}$ is the corresponding eigen mode amplitude. Substituting Eq.~(\ref{equ:eigenmode}) into Eqs.~(\ref{equ:H_t}) and (\ref{equ:DNS}), respectively, we obtain the Hamiltonian in the eigen mode basis and the evolution equation for the amplitude $c_{\alpha}$
\begin{align} 
&H=\sum_{\alpha}{\varepsilon _{\alpha}|c_{\alpha}|^2}+\frac{\chi}{2}\sum_{\alpha ,\beta ,\gamma ,\delta}{\Gamma _{\alpha \beta \gamma \delta}c_{\alpha}^{*}c_{\beta}^{*}c_{\gamma}c_{\delta}},
\label{equ:H_t2}
\\
&i\frac{dc_{\alpha}}{dz}=\varepsilon _{\alpha}c_{\alpha}+\chi \sum_{\beta ,\gamma ,\delta}{\Gamma _{\alpha \beta \gamma \delta}c_{\beta}^{*}c_{\gamma}c_{\delta}},
\label{equ:amplitudes}
\end{align}
where 
\begin{equation} 
\Gamma _{\alpha \beta \gamma \delta}=\sum_m{\phi _{\alpha}^{*}\left( m \right) \phi _{\beta}^{*}\left( m \right) \phi _{\gamma}\left( m \right) \phi _{\delta}\left( m \right)}
\end{equation}
is the four-wave mixing coefficient. 

Noting that, in the nonlinear summation of Eq.~(\ref{equ:H_t2}), the terms satisfying 
$\left( \alpha =\gamma ,\beta =\delta \right) $
 and 
$\left( \alpha =\delta ,\beta =\gamma \right) $ have phases that cancel each other, these contributions only produce a nonlinear frequency shift. In fact, they are proportional to $|c_{\beta}|^2c_{\alpha}$ and therefore merely renormalize the eigen frequency, $\varepsilon _{\alpha}\rightarrow \tilde{\varepsilon}_{\alpha}$, without changing the intensities $I_{\alpha}=|c_{\alpha}|^2$. As a result, these terms do not induce modal scattering and energy transfer, which arise instead from the remaining non‑secular terms. Thus, we decompose the Hamiltonian as $H=H_\text{mf}+H_{\text{int}}$, where $H_\text{mf}$ collects all terms that are diagonal in the modal intensities, including the nonlinear frequency shifts and $H_{\text{int}}$ contains the remaining interaction terms that induce mode coupling. It then follows that 
\begin{align} 
&H_\text{mf}=\sum_{\alpha}{\varepsilon _{\alpha}|c_{\alpha}|^2}+\chi \sum_{\alpha ,\beta}{\Gamma _{\alpha \beta \beta \alpha}|c_{\beta}|^2}|c_{\alpha}|^2,
\\
&H_{\text{int}}=\frac{\chi}{2}\sum_{\alpha ,\beta ,\gamma ,\delta}^{\prime}{\Gamma _{\alpha \beta \gamma \delta}c_{\alpha}^{*}c_{\beta}^{*}c_{\gamma}c_{\delta}},
\label{equ:H_int}
\end{align} 
where the prime in Eq.~(\ref{equ:H_int}) indicates that the sum runs over all terms except those 
 already included in $H_\text{mf}$ with $\left( \alpha =\gamma, \beta =\delta \right) $
 and 
$\left( \alpha =\delta, \beta =\gamma \right) $. The eigenvalues $\tilde{\varepsilon}_{\alpha}$ of $H_\text{mf}$ now include corrections due to nonlinear coupling
\begin{equation} 
\tilde{\varepsilon}_{\alpha}=\frac{\partial H_\text{mf}}{\partial \left| c_{\alpha} \right|^2}=\varepsilon_{\alpha}+2\chi \sum_{\beta}{\Gamma _{\alpha \beta \beta \alpha}|c_{\beta}|^2}.
\end{equation} 
We have used the tilde version to denote quantities with nonlinear correction. 

In the weakly nonlinear regime, Eq.~(\ref{equ:amplitudes}) can be treated perturbatively by substituting the solution of the linear equation, i.e., $c_{\alpha}=\sqrt{I_{\alpha}}\text{e}^{-i\omega _{\alpha}z-i\theta _{\alpha}^{0}}$, into the kinetic equation. This procedure can be carried out in two ways: one may use either the eigenvalues of the linear Hamiltonian $H_L$ (in which case $\omega _{\alpha}=\varepsilon _{\alpha}$) or those of the mean-field Hamiltonian $H_\text{mf}$ (in which case $\omega _{\alpha}=\tilde{\varepsilon}_{\alpha}$). In previous works \cite{shi_controlling_2021}, $H_L$ has been used, and the nonlinear frequency shift enters only through rapidly varying phases. After phase averaging, it does not appear explicitly in the resulting kinetic equation. Here, instead, we adopt $H_\text{mf}$ as the reference Hamiltonian. The kinetic equation then reads
\begin{equation} 
\frac{dI_{\alpha}}{dz}=\chi ^2\sum_{\beta \gamma \delta}^{\prime}{V_{\alpha \beta \gamma \delta}\left( I_{\beta}I_{\gamma}I_{\delta}+I_{\alpha}I_{\gamma}I_{\delta}-I_{\alpha}I_{\beta}I_{\delta}-I_{\alpha}I_{\beta}I_{\gamma} \right)},
\end{equation} 
where $V_{\alpha \beta \gamma \delta}$ is the scattering factor and expressed as
\begin{equation} 
V_{\alpha \beta \gamma \delta}=4\pi |\Gamma _{\alpha \beta \gamma \delta}|^2\delta \left( \tilde{\varepsilon}_{\alpha}+\tilde{\varepsilon}_{\beta}-\tilde{\varepsilon}_{\gamma}-\tilde{\varepsilon}_{\delta} \right) .
\label{equ:kinetic}
\end{equation} 
Here, the $\delta$-function enforces energy conservation in the scattering process, and the frequency corrections appear explicitly. It can be verified that when the system is in equilibrium, i.e., $dI_{\alpha}/dz=0$, Eq.~(5) is the solution to Eq.~(\ref{equ:kinetic}).

\section{S2. Optical Thermodynamic Theory} 
Starting from the mean-field Hamiltonian $H_\text{mf}$, the nonlinear theory follows the same line as the linear one\cite{wu_thermodynamic_2019}. 

\subsection{S2-A. Maximize optical entropy} 

Consider a continuous power cluster with total power $P$, distributed over $M$ distinct optical modes, each characterized by an energy level $\varepsilon_\alpha$  and a degeneracy $g_\alpha$. To describe the microscopic distribution of this power cluster, we decompose it into $N$ indistinguishable power packets (analogous to bosons), and assign $n_\alpha$ power packets to each energy level $\varepsilon_\alpha$. The corresponding number of microscopic states of the system is denoted by $\varOmega $,
\begin{equation} 
\varOmega =\prod_{\alpha}^M{\frac{\left( n_{\alpha}+g_{\alpha}-1 \right) !}{n_{\alpha}!\left( g_{\alpha}-1 \right) !}}.
\end{equation}
where $n_{\alpha}=n_c|c_{\alpha}|^2$, $|c_{\alpha}|^2$ is the normalized optical power in energy level $\varepsilon_\alpha$, and $n_c$ is the number of power packets contained in each unit of normalized optical power. 
Using the mean-field Hamiltonian, conservation of power packets number and energy implies
\begin{align} 
N&=\sum_{\alpha}^M{n_{\alpha}},
\label{equ:ys_P}
\\
E&=\sum_{\alpha}^M{\varepsilon _{\alpha}n_{\alpha}}+\frac{\chi}{n_c}\sum_{\alpha ,\beta}{\Gamma _{\alpha \beta \beta \alpha}n_{\beta}n_{\alpha}}.
\label{equ:ys_E}
\end{align} 

For the isolated system, we examine the maximization of the Boltzmann entropy $S_A=\ln \varOmega $, 
under the two constraints in Eqs.~(\ref{equ:ys_P}) and (\ref{equ:ys_E}). The distribution function can be obtained using the method of Lagrange multipliers, and takes the form of a Bose–Einstein distribution
\begin{align} 
&\frac{n_{\alpha}}{g_{\alpha}}=\left\{ \exp \left[ \alpha +\beta \left( \varepsilon _{\alpha}+2\frac{\chi}{n_c}\sum_{\alpha ,\beta}{\Gamma _{\alpha \beta \beta \alpha}n_{\beta}} \right) \right] -1 \right\} ^{-1}.
\end{align} 
In multimode nonlinear optical systems, the regime $n_{\alpha}\gg g_{\alpha}$ is often relevant, allowing the Bose–Einstein distribution to be approximated by the R–J distribution. 
\begin{equation} 
\frac{n_{\alpha}}{g_{\alpha}}=\left[ \alpha +\beta \left( \varepsilon _{\alpha}+2\frac{\chi}{n_c}\sum_{\alpha ,\beta}{\Gamma _{\alpha \beta \beta \alpha}n_{\beta}} \right) \right] ^{-1}
\label{equ:RJ_S}
\end{equation}

In thermodynamic theory, it is common to impose $\alpha \equiv -\tilde{\mu} /\left( n_cT \right) $ and $\beta \equiv 1/\left( n_cT \right) $. Taking $g_\alpha=1$, the modal occupancy Eq.~(\ref{equ:RJ_S}) can then be simplified to
\begin{equation} 
|c_{\alpha}|^2=\frac{T}{\varepsilon _{\alpha}+2\chi \sum_{\beta}{\Gamma _{\alpha \beta \beta \alpha}|c_{\beta}|^2}-\tilde{\mu}},
\end{equation}
where $T$ is the optical temperature, and $\tilde{\mu}$ is the chemical potential. The corresponding two conserved quantities are
\begin{align} 
P&=\sum_{\alpha}^M{|c_{\alpha}|^2},
\label{equ:Ps}
\\
\widetilde{U}&=\sum_{\alpha}{\varepsilon _{\alpha}|c_{\alpha}|^2}+\chi \sum_{\alpha ,\beta}{\Gamma _{\alpha \beta \beta \alpha}|c_{\beta}|^2}|c_{\alpha}|^2.
\label{equ:Us}
\end{align} 

Under the same conditions, the optical entropy is written as 
\begin{equation} 
S_A=\sum_{\alpha}^M{\ln n_{\alpha}}=\sum_{\alpha}^M{\ln \left| c_{\alpha} \right|^2}+M\ln n_c.
\label{equ:Sa_s}
\end{equation}
Further ignoring the constant term, one arrives at the final expression
\begin{equation} 
S=\sum_{\alpha}^M{\ln \left| c_{\alpha} \right|^2}.
\label{equ:Sr_s}
\end{equation}

\subsection{S2-B. Equation of state and extensivity} 

Using the expressions for optical power and internal energy
\begin{align}
P=\sum_{\alpha}^M{\frac{T}{\varepsilon _{\alpha}+2\chi \sum_{\beta}{\Gamma _{\alpha \beta \beta \alpha}|c_{\beta}|_{eq}^{2}}-\tilde{\mu}}},
\\
\widetilde{U}=\sum_{\alpha}^M{\frac{\left( \varepsilon _{\alpha}+\chi \sum_{\beta}{\Gamma _{\alpha \beta \beta \alpha}|c_{\beta}|_{eq}^{2}} \right) T}{\varepsilon _{\alpha}+2\chi \sum_{\beta}{\Gamma _{\alpha \beta \beta \alpha}|c_{\beta}|_{eq}^{2}}-\tilde{\mu}}},
\end{align}
we calculate $\widetilde{U}-\tilde{\mu} P$ and obtain
\begin{equation} 
\widetilde{U}-\tilde{\mu} P=\sum_{\alpha}^M{\frac{\left( \varepsilon _{\alpha}+\chi \sum_{\beta}{\Gamma _{\alpha \beta \beta \alpha}|c_{\beta}|_{eq}^{2}}-\tilde{\mu} \right) T}{\varepsilon _{\alpha}+2\chi \sum_{\beta}{\Gamma _{\alpha \beta \beta \alpha}|c_{\beta}|_{eq}^{2}}-\tilde{\mu}}}=MT-\chi \sum_{\alpha ,\beta}{\Gamma _{\alpha \beta \beta \alpha}|c_{\alpha}|_{eq}^{2}|c_{\beta}|_{eq}^{2}}.
\label{eq:ump}
\end{equation}
To obtain an EoS, we need to express the second term on the right side using macroscopic quantities. 
For common structures such as square, SSH, honeycomb, and triangular lattices under periodic boundary conditions, we have $\Gamma _{\alpha \beta \beta \alpha}=M^{-1}$. For generic structures, we adopt the average-intensity approximation which has been used in previous works, with $\left| c_{\beta} \right|^2\approx \left< \left| c_{\beta} \right|^2 \right> =P/M$\cite{PhysRevE.89.022921, PhysRevLett.132.193802}. Consequently, $\Gamma _{\alpha \beta \beta \alpha}$ decouples from the modal occupancies, leading to
\begin{equation} 
\sum_{\beta}{\Gamma _{\alpha \beta \beta \alpha}}=\sum_m{\phi _{\alpha}^{2}\left( m \right) \sum_{\beta}{\phi _{\beta}^{2}\left( m \right)}}=1,
\end{equation}
where we use the fact that the eigen mode matrix is unitary, so $\boldsymbol{\phi \phi }^{\dag}=\boldsymbol{\phi }^{\dag}\boldsymbol{\phi }=\boldsymbol{I}$.
Thus, we obtain the important approximation
\begin{equation} 
2\chi \sum_{\beta}{\Gamma _{\alpha \beta \beta \alpha}|c_{\beta}|_{eq}^{2}} = 2\chi {P}/{M}.
\end{equation}
Substituting into Eq.~(\ref{eq:ump}), the equation of state can then be obtained 
\begin{align}
\widetilde{U}-\left( \tilde{\mu} -\chi P/M \right) P=MT . 
\label{equ:state_s}
\end{align}

Consider scaling the system by a factor $\lambda $ while preserving its self-similarity (i.e., preserving the density of states profile), and simultaneously scaling the input by the same factor. With the extensive quantities $M\rightarrow \lambda M$, $P\rightarrow \lambda P$, $\widetilde{U}\rightarrow \lambda \widetilde{U}$, it can be observed that Eq.~(\ref{equ:state_s}) still applies. For the optical entropy $S\left(  M,P, \widetilde{U} \right)$, assuming that after scaling the system, the split energy levels remain very close to the original level $\varepsilon _{\alpha}$, then we have
\begin{equation} 
S\left( \lambda M,\lambda P,\lambda \widetilde{U} \right) =\sum_{\alpha}^{\lambda M}{\frac{T}{\varepsilon _{\alpha}+2\chi \frac{\lambda P}{\lambda M}-\tilde{\mu}}}\approx \lambda \sum_{\alpha}^M{\frac{T}{\varepsilon _{\alpha}+2\chi \frac{P}{M}-\tilde{\mu}}}=\lambda S\left(  M,P, \widetilde{U} \right).
\label{equ:ext_s}
\end{equation}

\subsection{S2-C. Fundamental thermodynamic relation} 
We show in this section that, similar to the linear case, the following three relations hold
\begin{align}
 \left( \frac{\partial S}{\partial \widetilde{U}} \right) _{P,M}=\frac{1}{T}, \quad 
\left( \frac{\partial S}{\partial P} \right) _{\widetilde{U},M}=-\frac{\tilde{\mu}}{T}, \quad
\left( \frac{\partial S}{\partial M} \right) _{\widetilde{U},P}=\frac{\tilde{p}}{T}.
\label{equ:pS_pU}
\end{align}
First, from the definition of entropy and the EoS, we have
\begin{equation} 
S=\sum_{\alpha}^M{\ln \frac{T}{\varepsilon _{\alpha}+2\chi P/M-\tilde{\mu}}}=\sum_{\alpha}^M{\ln \frac{T}{\varepsilon _{\alpha}+\chi P/M-\frac{\widetilde{U}-MT}{P}}}.
\end{equation}
For notational simplicity, we omit the superscript $(M)$ on $\chi ^{\left( M \right)}$. It then follows that 
\begin{equation} 
\left( \frac{\partial S}{\partial \widetilde{U}} \right) _{P,M}=\left( \frac{\partial S}{\partial \widetilde{U}} \right) _T+\left( \frac{\partial S}{\partial T} \right) _{\widetilde{U}}\left( \frac{\partial T}{\partial \widetilde{U}} \right) _{P,M}.
\end{equation}
For $\left( \partial S/\partial T \right) _{\widetilde{U}}$, we have
\begin{equation} 
\left( \frac{\partial S}{\partial T} \right) _{\widetilde{U}}=\sum_{\alpha}^M{\frac{1}{T}}-\sum_{\alpha}^M{\frac{M/P}{\varepsilon _{\alpha}+\chi P/M-\frac{\widetilde{U}-MT}{P}}}=\frac{M}{T}-\frac{M}{P}\frac{P}{T}=0, 
\end{equation}
while for $\left( \partial S/\partial \widetilde{U} \right) _T$, we have
\begin{equation} 
\left( \frac{\partial S}{\partial \widetilde{U}} \right) _T=-\sum_{\alpha}^M{\frac{-1/P}{\varepsilon _{\alpha}+\chi P/M-\frac{\widetilde{U}-MT}{P}}}=\frac{1}{P}\frac{P}{T}=\frac{1}{T}.
\end{equation}
Therefore, we arrive at
\[ 
\left( \frac{\partial S}{\partial \widetilde{U}} \right) _{P,M}=\frac{1}{T} .
\]

Similarly, using the EoS, the entropy is written  as
\begin{align} 
S=\sum_{\alpha}^M{\ln \frac{\widetilde{U}-\tilde{\mu} P+\chi P^2/M}{\left( \varepsilon _{\alpha}+2\chi P/M-\tilde{\mu} \right) M}}.
\end{align}
Writing
\begin{align}
\left( \frac{\partial S}{\partial P} \right) _{\widetilde{U},M}=\left( \frac{\partial S}{\partial P} \right) _{\tilde{\mu}}+\left( \frac{\partial S}{\partial \tilde{\mu}} \right) _P\left( \frac{\partial \tilde{\mu}}{\partial P} \right) _{\widetilde{U},M},
\end{align} 
following the same steps, it can be shown that
\begin{align}
\left( \frac{\partial S}{\partial P} \right) _{\widetilde{U},M}=-\frac{\tilde{\mu}}{T}.
\end{align}

Since $\varepsilon$ also depends on $M$, we introduce the density of states $D\left( \varepsilon \right) =dM/d\varepsilon $ and the volume $V=M/M_0$, where $M_0$ denotes the number of modes in the reference system. We consider only the self-similar extension case, and the number of modes for the reference system is $\int{D\left( \varepsilon \right)}d\varepsilon =M_0$. When $M_0$ is large enough, transforming the summation into an integral to obtain
\begin{equation} 
S=\int{VD\left( \varepsilon \right) \ln \frac{T}{\varepsilon _{\alpha}+2\chi P/M-\tilde{\mu}}}d\varepsilon =\int{VD\left( \varepsilon \right) \ln \frac{T}{\varepsilon _{\alpha}+\chi P/M-\frac{\widetilde{U}-MT}{P}}}d\varepsilon.
\end{equation}
Starting from   
\begin{equation} 
\left( \frac{\partial S}{\partial M} \right) _{\widetilde{U},P}=\left( \frac{\partial S}{\partial M} \right) _T+\left( \frac{\partial S}{\partial T} \right) _M\left( \frac{\partial T}{\partial M} \right) _{\widetilde{U},P},
\end{equation}
it can be shown that 
\begin{equation} 
\left( \frac{\partial S}{\partial T} \right) _M = 0, \quad 
\left( \frac{\partial S}{\partial M} \right) _T=\frac{S}{M}-1+\chi \frac{P^2}{TM^2}.
\end{equation}
%
%
Defining the optical thermodynamic pressure $\tilde{p}$ as 
\begin{equation} 
\tilde{p} \equiv T\left( \frac{S}{M}-1 \right) +\chi \frac{P^2}{M^2},
\label{equ:photo_p}
\end{equation}
one arrives at 
\[
\left( \frac{\partial S}{\partial M} \right) _{\widetilde{U},P}=\frac{\tilde{p}}{T}.
\]
The first term on the right-hand side of Eq.~(\ref{equ:photo_p}) is consistent with linear theory, while the second term represents the nonlinear correction, similar to the pressure correction form in the van der Waals equation. Being an intensive quantity, $\tilde{p}$ satisfies $\tilde{p}\left( \lambda M,\lambda P,\lambda \widetilde{U} \right) =\tilde{p}\left( M,P,\widetilde{U} \right) $, as follows from its definition Eq.~(\ref{equ:photo_p}) and the extensivity of entropy Eq.~(\ref{equ:ext_s}). 


\section{S3. Determination of the Temperature and the Chemical Potential} 

Given an initial input $\left| c_{\alpha} \right|^2$, the optical power $P$ and internal energy $\widetilde{U}$ of the optical array can be determined via Eqs.~(\ref{equ:Ps}) and (\ref{equ:Us}). To determine $T$ and $\tilde{\mu}$, we follow Ref.~\onlinecite{parto_thermodynamic_2019}. 
At equilibrium, using the EoS, the optical power is expressed as
\begin{equation} 
P=\sum_{\alpha}^M{\frac{T}{\varepsilon _{\alpha}+\chi P/M-(\widetilde{U}-MT)/P}}=\sum_{\alpha}^M{\frac{T}{\varepsilon _{\alpha}-(U_L-MT)/{P}}}.
\label{equ:S1}
\end{equation}
The second equality is exactly the same as the linear theory, which ensures that the linear and nonlinear theory gives the same temperature.
%
After determining $T$, the chemical potential is obtained from the EoS. The nonlinear results $\tilde{\mu}$ acquires an additional correction, compared to the linear result $\mu$:
\begin{equation} 
\left. \begin{array}{r}
	U_L+\frac{\chi P^2}{M}-\left( \tilde{\mu} -\frac{\chi P}{M} \right) P=MT\\
	U_L-\mu P=MT\\
\end{array} \right\} \ \ \Rightarrow \ \ \tilde{\mu}=\mu +2\chi \frac{P}{M}.
\label{equ:S_mu}
\end{equation}

For junctions as shown in Fig.~\ref{fig:S1}(a), we neglect the contribution of coupling between reservoirs to the total energy and power, $\widetilde{U}\approx \widetilde{U}_A+\widetilde{U}_B$ and $P=P_A+P_B$. The set $(T, \tilde{\mu})$ is obtained numerically by minimizing the criterion $\Delta R$ defined as
\begin{equation} 
\Delta R=\left |\frac{P-P_c}{P}\right |+\left |\frac{\widetilde{U}-\widetilde{U}_c}{\widetilde{U}}\right |, 
\end{equation}
where $P_c$ and $\widetilde{U}_c$ represent the optical power and internal energy corresponding to each $(T, \tilde{\mu})$ in the parameter space.

\section{S4. Thermalization between two systems with different nonlinear coefficients} 

For an isolated system, when two subsystems are in contact and allowed to exchange energy and power, the second law of thermodynamics requires that, in the equilibrium state, they attain the same temperature and chemical potential.  However, the linear theory gives unequal chemical potentials, violating thermodynamic expectation. The discrepancy depends on the nonlinear coefficient and the power density, and it is resolved by the nonlinear theory.

\begin{figure} [h]
\centering
\includegraphics[width=0.9\textwidth]{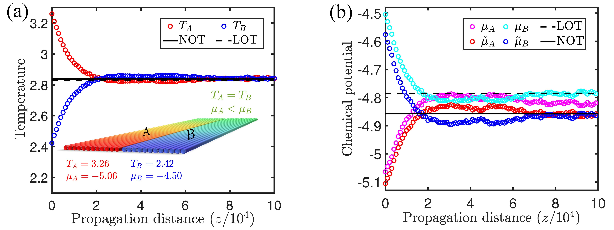}
\caption{Thermalization of two subsystems with different nonlinear coefficients. Parameters:  Number of modes $M_A=M_B=100$,  nonlinear coefficients $\chi_A=-0.03$, $\chi_B=-0.06$, nearest neighbor coupling $\kappa_A=\kappa_B=1$, subsystem coupling $\kappa_{AB}=0.1$.
(a) The two subsystems ($A$ and $B$) are initially in their respective equilibrium states with parameters shown in the inset. After turning on the coupling, the composite system evolves to the same temperature. The predictions of linear (LOT) and nonlinear (NOT) theories agree with each other. 
(b) The chemical potentials obtained from the linear theory  ($\mu_{A/B}$) are different from each other even when the system reaches equilibrium, while those obtained from the nonlinear theory ($\tilde{\mu}_{A/B}$) reach the same value.  
}
     \label{fig:S1}
\end{figure}

Consider a 1D waveguide array  consisting of two regions, A and B, with different nonlinear coefficients [Fig.~\ref{fig:S1}(a), inset]. Initially, each region is prepared in a different thermal equilibrium state. When the coupling $\kappa_\text{AB}$
between them is turned on, the combined system evolves toward thermodynamic equilibrium. The corresponding change in entropy $dS$ is given by
\begin{equation} 
dS=\left( -\frac{1}{T_A}+\frac{1}{T_B} \right) dU+\left( \frac{\mu _A}{T_A}-\frac{\mu _B}{T_B} \right) dP\ge 0.
\label{equ:S_S1}
\end{equation}
At thermal equilibrium, the entropy is maximal, so $dS=0$, which implies $T_A=T_B$ and $\mu_A=\mu_B$.

We solve the DNSE numerically and extract the temperature and chemical potential using both linear and nonlinear theories. Upon reaching equilibrium, the two theories yield the same temperature for the two subsystems [Fig.~\ref{fig:S1}(a)]. However, the linear theory gives unequal chemical potentials [Fig.~\ref{fig:S1}(b)], in contradiction with the prediction of Eq.~(\ref{equ:S_S1}). This inconsistency is resolved by the nonlinear theory, 
which gives equal chemical potentials  at thermal equilibrium $\tilde{\mu}_A=\tilde{\mu}_B$ [Fig.~\ref{fig:S1}(b)]. 

\section{S5. isothermal power elastic modulus \texorpdfstring{$\kappa^P_T$}{kappa\^{}P\_T}} 

The isothermal power elastic modulus is defined as 
\begin{align}\kappa^P _T=P\left( \frac{\partial \tilde{p}}{\partial P} \right) _{T,M} .
\end{align}  
Using Eq.~(10), we have
\begin{equation} 
\kappa^P _T=\frac{PT}{M}\left( \frac{\partial S}{\partial P} \right) _{T,M}+2\chi \frac{P^2}{M^2}.
\label{equ:k_s1}
\end{equation}
From the total differential of Helmholtz free energy $dF=-SdT-\tilde{p}dM+\tilde{\mu} dP$, we obtain the Maxwell relation
\begin{equation} 
\left( \frac{\partial S}{\partial P} \right) _{T,M}=-\left( \frac{\partial \tilde{\mu}}{\partial T} \right) _{P,M}.
\end{equation}
Substituting into Eq.~(\ref{equ:k_s1}) gives
\begin{equation} 
\kappa^P _T=-\frac{PT}{M}\left( \frac{\partial \tilde{\mu}}{\partial T} \right) _{P,M}+2\chi \frac{P^2}{M^2}.
\label{equ:kappa_s}
\end{equation}

Since $\tilde{p}$ is an intensive quantity, taking $T$, $M$, and $P$ as independent variables leads to $\tilde{p}\left( T,\lambda M,\lambda P \right) =\tilde{p}\left( T,M,P \right) $. Taking the derivative with respect to $\lambda$ and setting $\lambda=1$, we obtain
\begin{equation} 
P\left( \frac{\partial \tilde{p}}{\partial P} \right) _{T,M}=-M\left( \frac{\partial \tilde{p}}{\partial M} \right) _{T,P}\ \ \Rightarrow \ \ \kappa _{T}^{P}=\kappa _{T}^{M},
\end{equation}
where we have defined the isothermal volumetric elastic modulus
\begin{align}
\kappa _{T}^{M}=-M\left( \frac{\partial \tilde{p}}{\partial M} \right) _{T,P} .
\end{align}

We now give the specific expression for $\kappa^P _T$. The following two equations can be derived from Eqs.~(7) and (8) 
\begin{equation} 
\left( \frac{\partial \tilde{\mu}}{\partial T} \right) _{P,M}=\frac{1}{P}\left( \frac{\partial \widetilde{U}}{\partial T} \right) _{P,M}-\frac{M}{P},
\label{equ:mu_T_s}
\end{equation}
\begin{equation} 
\left( \frac{\partial \widetilde{U}}{\partial T} \right) _{P,M}=\frac{U_L}{T}-\frac{\varLambda}{P}\left( \frac{\partial \widetilde{U}}{\partial T} \right) _{P,M}+\frac{M\varLambda}{P},
\label{equ:U_T_s}
\end{equation}
where we have defined 
\begin{equation} 
\varLambda  \equiv \sum_{\alpha}^M{\varepsilon _{\alpha}\frac{\partial |c_{\alpha}|^2}{\partial \varepsilon _{\alpha}}}=-\sum_{\alpha}^M{\frac{T\varepsilon _{\alpha}}{\left( \varepsilon _{\alpha}+\chi P/M-\frac{\widetilde{U}-MT}{P} \right) ^2}}.
\end{equation}
Thus, by Eq.~(\ref{equ:U_T_s}), we have
\begin{equation} 
\left( \frac{\partial \widetilde{U}}{\partial T} \right) _{P,M}=\frac{P}{P+\varLambda}\left( \frac{U_L}{T}+\frac{M\varLambda}{P} \right) . 
\label{equ:u-t-s}
\end{equation}
According to Eqs.~(\ref{equ:kappa_s}) and (\ref{equ:mu_T_s}), it follows that $\kappa^P _T$ satisfies
\begin{equation} 
\kappa _{T}^{P}=\frac{P}{P+\varLambda}\left( T-\frac{U_L}{M} \right) +2\chi \frac{P^2}{M^2}.
\end{equation}

Next, we determine the sign of $\kappa^P _T$ when neglecting the nonlinearity. Equation~(\ref{equ:kappa_s}) indicates that it is determined by $\left( \partial {\mu} /\partial T \right) _{P,M}$. For an equilibrium system with constant $P$ and $M$, let $T$ increases by $dT$ (with $dT > 0$). For $T>0$, an increase in $T$ leads to an increase in the optical power $P$, and the $\mu$ must move further away from the lowest energy level $\varepsilon _1$ to keep $P$ constant.  Conversely, for $T < 0$, $P$ decreases, and to keep $P$ constant in this case, $\mu$ must move closer to the highest energy level $\varepsilon _M$. In either case, the result is $d {\mu} < 0$, i.e., $\left( \partial {\mu} /\partial T \right) _{P,M}<0$. Therefore, we have $\kappa^P _T \lessgtr 0$ when $T \lessgtr 0$.

\section{S6. Thermodynamic stability analysis of the system} 

Consider a local optical power fluctuation $(\delta P_A>0)$ in region A of the waveguide array. Due to optical power conservation, the power in the remaining region B decreases by $\delta P_A$. The corresponding change in the local pressure is
\begin{equation} 
\begin{aligned}
&\delta \tilde{p}_A=\left( \frac{\partial \tilde{p}}{\partial P} \right) _{T,M}\delta P_A=\kappa _{T}^{P}\frac{\delta P_A}{P}, \\
&\delta \tilde{p}_B=\left( \frac{\partial \tilde{p}}{\partial P} \right) _{T,M}\left( -\delta P_A \right) =\kappa _{T}^{P}\left( -\frac{\delta P_A}{P} \right) .
\end{aligned}
\end{equation}
When $\kappa _{T}^{P}>0$,  the pressure in region A increases while that in region B decreases, whereas for $\kappa^P_T<0$ the situation is reversed. 

Next, we analyze the direction of optical power transfer. Assume that regions A and B are initially separated by a partition, and their volumes are $M_A$ and $M_B$, respectively. According to the second law of thermodynamics, we have
\begin{equation} 
dS=\left( -\frac{\tilde{p}_A}{T_A}+\frac{\tilde{p}_B}{T_B} \right) dM_B\ge 0.
\end{equation}
When the two regions have the same temperature and $\tilde{p}_A>\tilde{p}_B$, for positive temperature the entropy increase requires $dM_B<0$, i.e., region A tends to expand; once the partition is removed, optical power  transfers from the high-pressure region to the low-pressure region.  For negative temperature, the entropy increase requires $dM_B>0$, i.e., region A tends to shrink, and after the partition is removed, optical power transfers from the low-pressure region to the high-pressure region.

Thus, for $T>0, \kappa^P_T>0$ (or $T<0, \kappa^P_T<0$), the additional power $\delta P_A$ in region A is transferred to region B, which is at lower (or higher) pressure; the fluctuation is therefore suppressed and the system remains stable. In contrast, for $T>0, \kappa^P_T<0$ (or $T<0, \kappa^P_T>0$), power in region B, which is at higher (or lower) pressure, is continuously transferred into region A, so that the fluctuation is amplified and the system becomes unstable.

\section{S7. Optical Joule–Thomson expansion coefficient}
We use the expansion coefficient
\begin{align}
\eta=\left( \frac{\partial T}{\partial M} \right) _{P,\widetilde{U}}
\end{align}
to characterize the temperature change resulting from change of $M$ under a given input $(P,\widetilde{U})$. Writing $\beta =T^{-1}$, 
it is further written as 
\begin{equation} 
\eta =-T^2\left( \frac{\partial \beta}{\partial M} \right) _{P,\widetilde{U}}.
\label{equ:JT1_s}
\end{equation} 
Through the optical entropy $dS=\beta d\widetilde{U}+\beta \tilde{p}dM-\beta \tilde{\mu} dP$, we obtain the Maxwell relation
\begin{equation} 
\left( \frac{\partial \beta}{\partial M} \right) _{P,\widetilde{U}}=\left( \frac{\partial \left( \beta \tilde{p} \right)}{\partial \widetilde{U}} \right) _{P,M}. 
\label{equ:JT2_s}
\end{equation} 
Using Eqs.~(10) and (\ref{equ:pS_pU}), the expansion coefficient can be rewritten as 
\begin{align} 
\eta &=-\frac{T}{M}+\chi \frac{P^2}{M^2}\left( \frac{\partial T}{\partial {U_L}} \right) _{P,M}\\
&=-\frac{T}{M}+\chi \frac{P^2}{M^2}\frac{PT+\varLambda T}{PU_L+MT\varLambda}.
\end{align}
Eq.~(\ref{equ:u-t-s}) has been used to obtain the second line.

Next, we analyze the sign of $\eta$. For a system with $\bar{\varepsilon}=0$ ($\bar{\varepsilon}$ denotes the average eigenvalue of $H_L$), the sign of $T$ is determined by $U_L$. As shown in Fig.~\ref{fig:T_U_s}(a), $U_L>0$ gives $T<0$, while $U_L<0$ gives $T>0$, and $\left( \partial T/\partial U_L \right) _{P,M}>0$.  Therefore, as shown in Fig.~\ref{fig:T_U_s}(b), for $T>0$ and $\chi<0$ we have $\eta<0$, so increasing $M$ cools the system; whereas for $T<0$ and $\chi>0$ we have $\eta>0$, so increasing $M$ heats the system. When $T$ and $\chi$ have the same sign, as described in the main text, $\eta$ is determined by the competition between $U_L$ and $U_{NL}$. In Fig.~\ref{fig:T_U_s}(c), we present the results for $T<0$ and $\chi<0$, which are symmetric to the parameters in Figs.~3(d) and 3(f) of the main text.


\clearpage

\section{Additional figures} 

\begin{figure}[h]
\centering
 \includegraphics[width=0.9\textwidth]{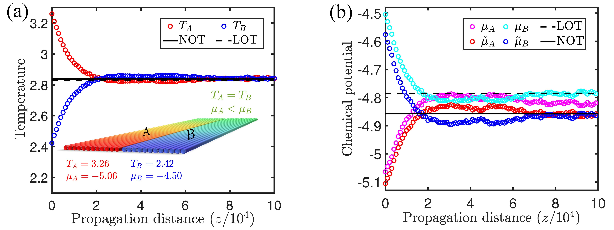}
\caption{Instability of a uniform 1D waveguide array shown in Fig.~2(a) at negative temperature $T=-0.5$. All other parameters are the same as Fig.~2. (a) The isothermal power elastic modulus $\kappa^P_T$. (b) Evolution of the optical power $|\psi _m|^2$ along $z$ at $P/M=$0.5 (I), 0.8 (II), and 1.3 (III, IV), with one ensemble randomly selected for each case. (c) The corresponding complex-$\psi$ space obtained from 500 ensembles, with the color indicating optical power $|\psi _m|^2$. 
When $\chi>0$, the high input power drives the system unstable and leads to localization, whereas for $\chi<0$ the system remains stable. 
}
\label{fig:fwd_S}
\end{figure}

\begin{figure}[h]
\centering
   \includegraphics[width=\textwidth]{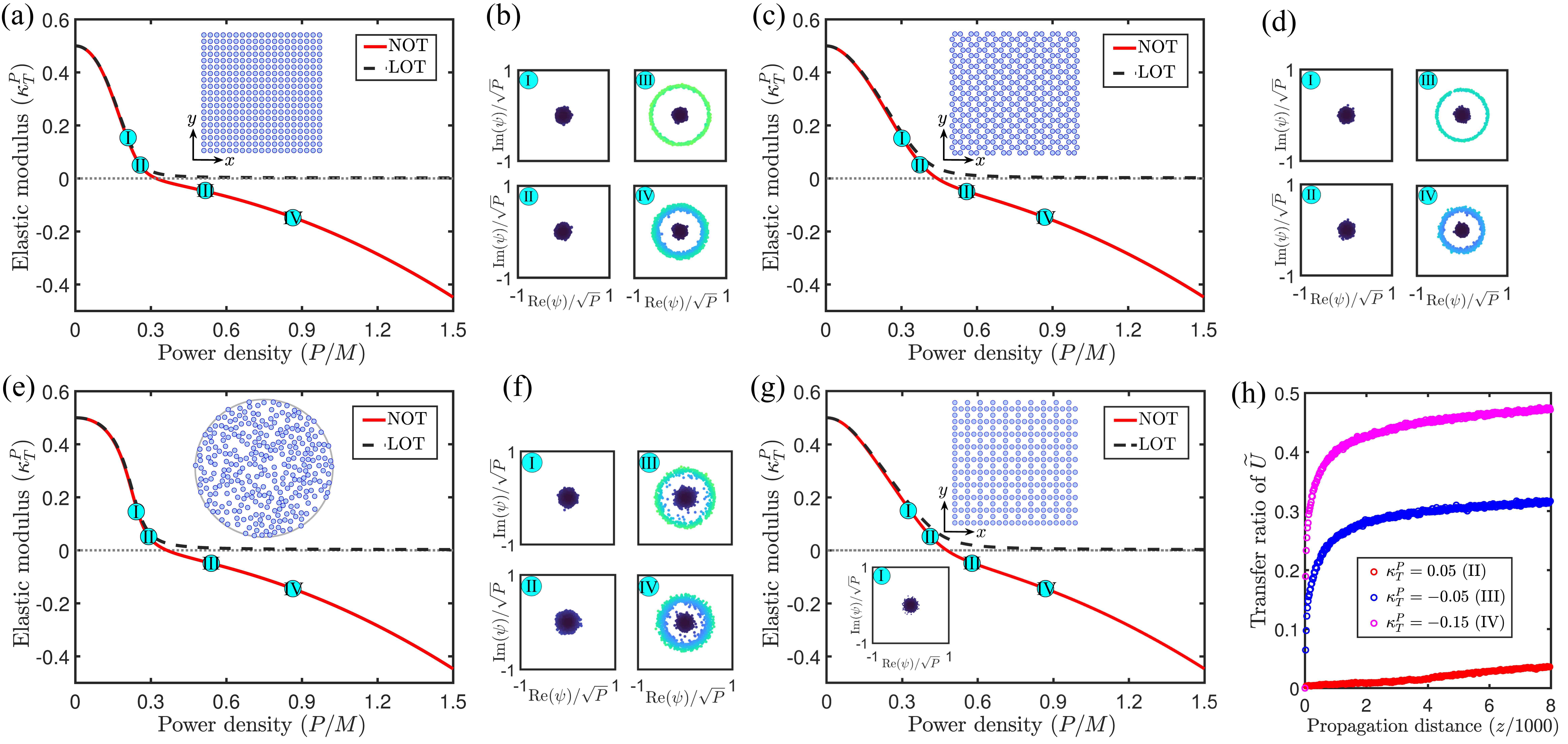}
\caption{The isothermal power elastic modulus $\kappa^P_T$ for (a) square ($M=M_x \times M_y =19 \times 19$), (c) honeycomb ($M=M_x \times M_y =16 \times 23$), (e) irregular ($M=300$) and (f) Lieb array ($M=3\times N_x \times N_y =3\times10 \times 10$) with $\kappa=1$, $\chi=0.1$ at $T=0.5$. 
For the irregular array, 300 sites were randomly generated within a circular region of radius 15, with the inter-site spacing $l_{mn}\ge 1$. The coupling coefficient is set to $\kappa _{mn}=\exp \left[ -2\left( l_{mn}-1 \right) \right] $. 
(b, d, f) The complex-$\psi$ space distribution obtained from 500 ensembles for $\kappa^P_T$=0.15 (I), 0.05 (II), -0.05 (III), and -0.15 (IV). (h) Ratio of energy transferred to the nonlinear part $1-\widetilde{U}(z)/\widetilde{U}(0)$ as a function of propagation distance for the Lieb array.
All results are consistent with the theoretical predictions.
When $\kappa^P_T=0.15$, the system thermalizes as described. When $\kappa^P_T=0.05$, the energy gradually transfers from the linear to the nonlinear part, indicating that nonlinearity starts to compete with diffraction. For $\kappa^P_T<0$, the optical power becomes strongly localized.
}
\label{fig:square_Liujiao_S}
\end{figure}

\begin{figure}[h]
\centering
  \includegraphics[width=0.9\textwidth]{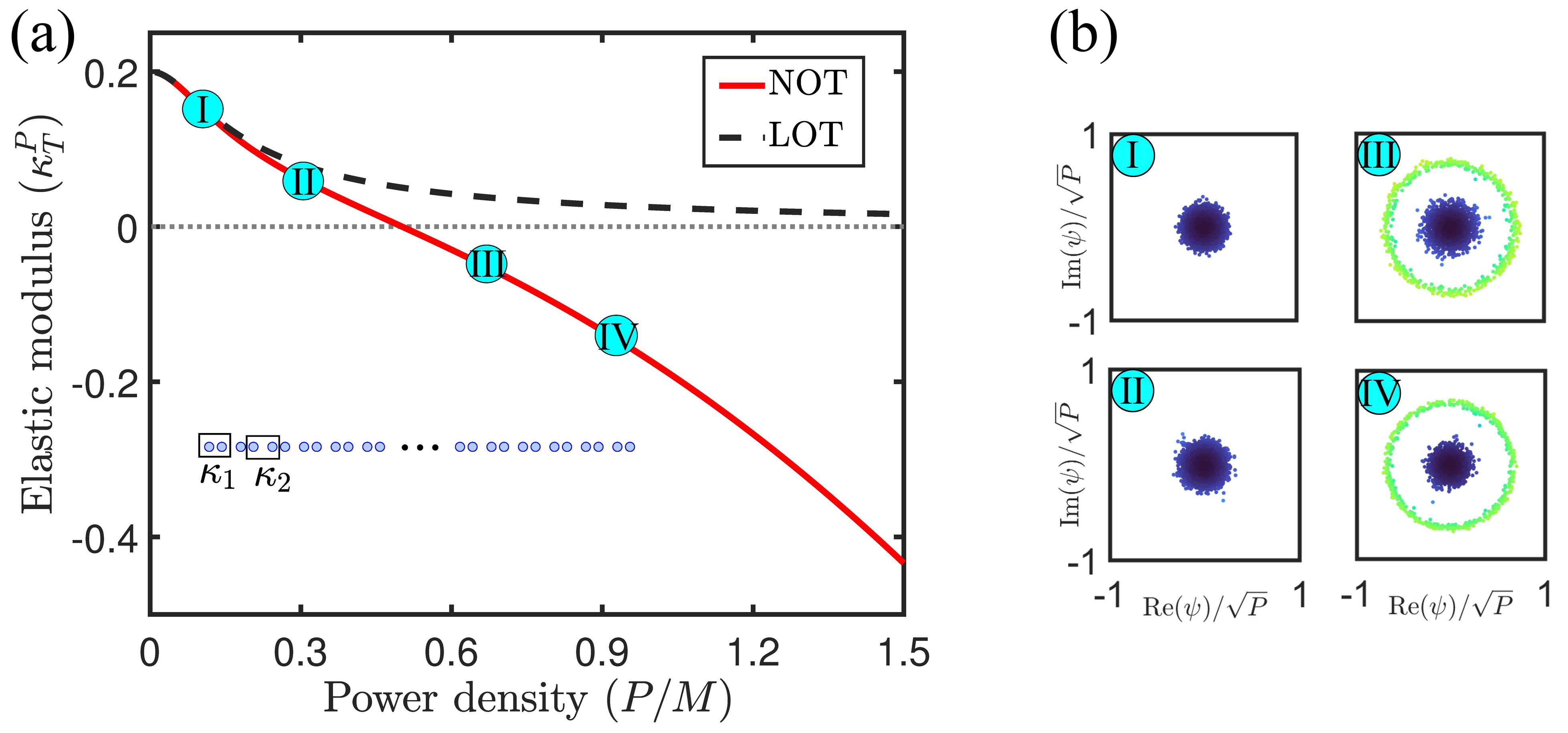}
\caption{(a) The isothermal power elastic modulus $\kappa^P_T$ for SSH array. Parameters: $M = 100$, $\kappa_1 = 1$, $\kappa_2 = 0.6$, $\chi = 0.1$, $T = 0.2$. (b) The corresponding complex-$\psi$ space distribution for $\kappa^P_T$=0.15 (I), 0.05 (II), -0.05 (III), and -0.15 (IV).
}
\label{fig:SSH_S}
\end{figure}

\begin{figure}[h]
\centering
   \includegraphics[width=0.9\textwidth]{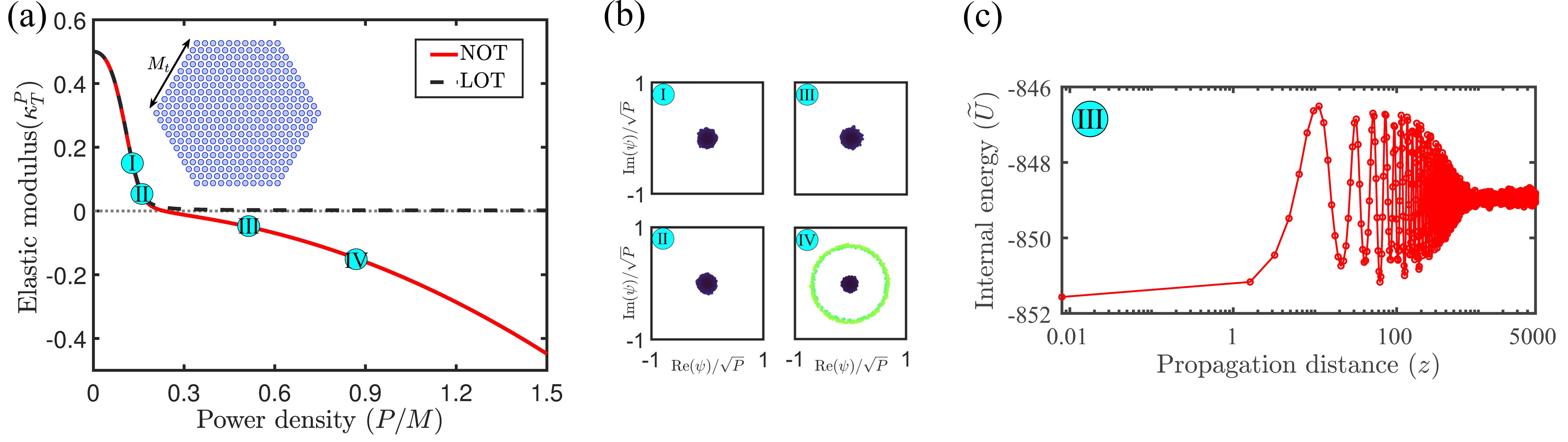}
\caption{(a) The isothermal power elastic modulus $\kappa^P_T$ for triangular waveguide array. Parameters: $M_t = 11$, $M = 331$, $\kappa = 1$,  $\chi = 0.1$, $T = 0.5$. (b) The corresponding complex-$\psi$ space distribution for $\kappa^P_T$=0.15 (I), 0.05 (II), -0.05 (III), and -0.15 (IV). (c) Evolution of internal energy $\widetilde{U}$ at $\kappa^P_T=-0.05$, showing that localization is more difficult to take place, due to a larger number of nearest neighbors.} 
\label{fig:Sanjiao_S}
\end{figure}

\begin{figure}[h]
\centering
  \includegraphics[width=0.9\textwidth]{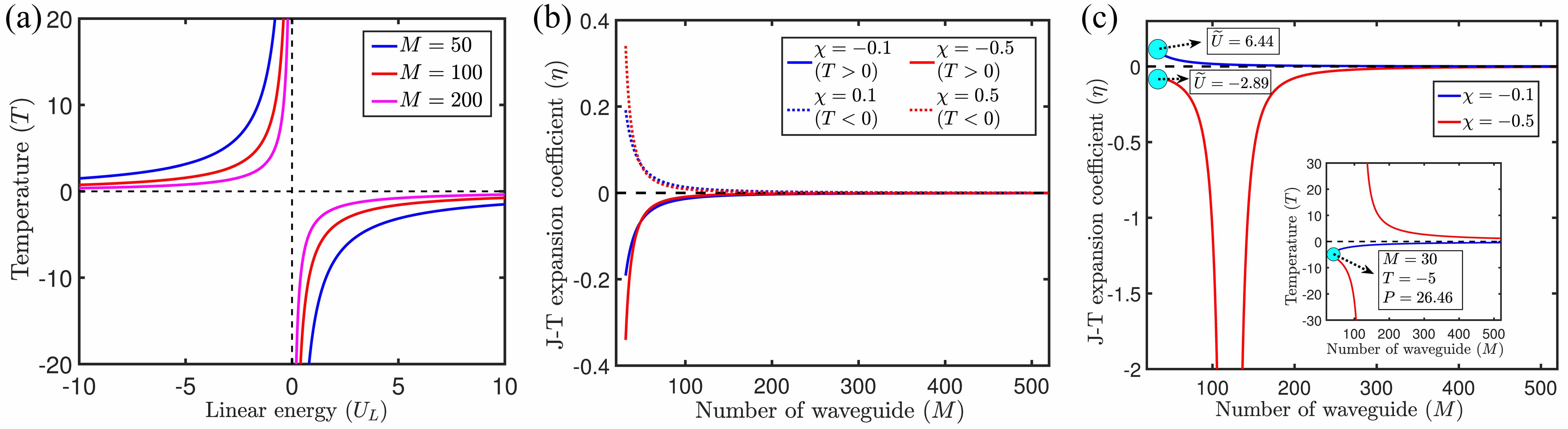}
\caption{(a) $U_L-T$ relationship of the 1D homogenous waveguide array. (b) The optical J–T response coefficient $\eta$ for opposite signs of $T$ and $\chi$. Initial condition: $T=\pm 5$, $P=26.46$, and $M=30$. (c) $\eta$ for $T<0$ and $\chi<0$. The behavior is opposite to the case of Fig.~3(d).
}
\label{fig:T_U_s}
\end{figure}

\end{document}